# Large Thermoelectric Power Factor in One-Dimensional Telluride $Nb_4SiTe_4$ and Substituted Compounds

Yoshihiko Okamoto,[1,3,a)] Taichi Wada,[1] Youichi Yamakawa,[2,3] Takumi Inohara,[1] and Koshi Takenaka[1]

[1]*Department of Applied Physics, Nagoya University, Nagoya 464-8603, Japan*
[2]*Department of Physics, Nagoya University, Nagoya 464-8602, Japan*
[3]*Institute for Advanced Research, Nagoya University, Nagoya 464-8601, Japan*

We found that whisker crystals of Mo-doped $Nb_4SiTe_4$ show high thermoelectric performances at low temperatures, indicated by the largest power factor of ~70 μW $cm^{-1}$ $K^{-2}$ at 230-300 K, much larger than those of $Bi_2Te_3$-based practical materials. This power factor is smaller than the maximum value in the 5*d* analogue $Ta_4SiTe_4$, but is comparable to that with a similar doping level. First principles calculation results suggest that the difference in thermoelectric performances between Nb and Ta compounds is caused by the much smaller band gap in $Nb_4SiTe_4$ than that in $Ta_4SiTe_4$, due to the weaker spin-orbit coupling in the former. We also demonstrated that the solid solution of $Nb_4SiTe_4$ and $Ta_4SiTe_4$ shows a large power factor, indicating that their combination is promising as a practical thermoelectric material, as in the case of $Bi_2Te_3$ and $Sb_2Te_3$. These results advance our understanding of the mechanism of high thermoelectric performances in this one-dimensional telluride system, as well as indicating the high potential of this system as a practical thermoelectric material for low temperature applications.

Recently, whisker crystals of the one-dimensional telluride $Ta_4SiTe_4$ and its substituted compounds were found to show high thermoelectric performances at low temperatures.[1] Their whisker form is typically several mm long and several μm in diameter, reflecting the strongly one-dimensional crystal structure comprising $Ta_4SiTe_4$ chains.[2,3] The electrical resistivity ρ and thermoelectric power *S* data measured along the whiskers, i.e. parallel to the $Ta_4SiTe_4$ chains, indicate that the power factor $P = S^2/\rho$ of the chemically-doped $Ta_4SiTe_4$ whiskers are significantly larger than those of practical thermoelectric materials in a wide temperature region of 50–300 K.[1] Undoped $Ta_4SiTe_4$ shows a very large and negative thermoelectric power of $|S|$ ~ 400 μV $K^{-1}$ at 100–200 K, while maintaining a small ρ of ~2 mΩ cm. These $|S|$ and ρ yield $P$ = 80 μW $cm^{-1}$ $K^{-2}$ at the optimum temperature of ~130 K, which is almost twice as large as those of $Bi_2Te_3$-based materials at room temperature. This *P* is strongly enhanced by electron doping. $(Ta_{1-x}Mo_x)_4SiTe_4$ with $x$ = 0.001–0.002 shows $P$ = 170 μW $cm^{-1}$ $K^{-2}$ at 220–280 K. These results indicate that $Ta_4SiTe_4$ is promising for low temperature applications of thermoelectric conversion, which have never been put to practical use, such as local cooling of electronic devices well below room temperature and power generation utilizing the cold heat of liquefied natural gas.

For the practical use of $Ta_4SiTe_4$ as a low-temperature thermoelectric material, it is necessary to study the thermoelectric properties of a sister compound of $Ta_4SiTe_4$ with the same crystal structure and electron configuration. Comparing the thermoelectric properties of $Ta_4SiTe_4$ to those of a sister compound will help us to understand the physics underlying the high thermoelectric performance in this system. Moreover, forming a solid solution or superlattice structure with sister compounds can improve the thermoelectric performances. In the $Bi_2Te_3$-based materials, for example, the thermoelectric performance is optimized by controlling the electronic state by forming a solid solution with the sister compounds $Sb_2Te_3$ and $Bi_2Se_3$. A $Bi_2Te_3/Sb_2Te_3$ superlattice thin film was reported to show a very large dimensionless figure of merit $ZT$ = 2.4.[4]

In this letter, we focus on $Nb_4SiTe_4$ as a sister compound of $Ta_4SiTe_4$. $Nb_4SiTe_4$ was first synthesized by Badding *et al.* and was reported to have the same crystal structure as that of $Ta_4SiTe_4$, as shown in Fig. 1(a), which is orthorhombic with the space group *Pbam*.[5] The ρ of a $Nb_4SiTe_4$ single crystal along the chain direction was reported to be ~3 mΩ cm at room temperature.[6] The ρ decreases with decreasing temperature from 260 to 50 K and exhibits a local minimum of 1.5 mΩ cm at ~50 K.[1,5,6] We prepared a series of whisker crystals of chemically doped $Nb_4SiTe_4$ and measured their ρ and *S*. The Mo-doped whisker crystals were found to show the largest *P* of ~70 μW $cm^{-1}$ $K^{-2}$ at 230-300 K, far exceeding

a) Electronic mail: yokamoto@nuap.nagoya-u.ac.jp

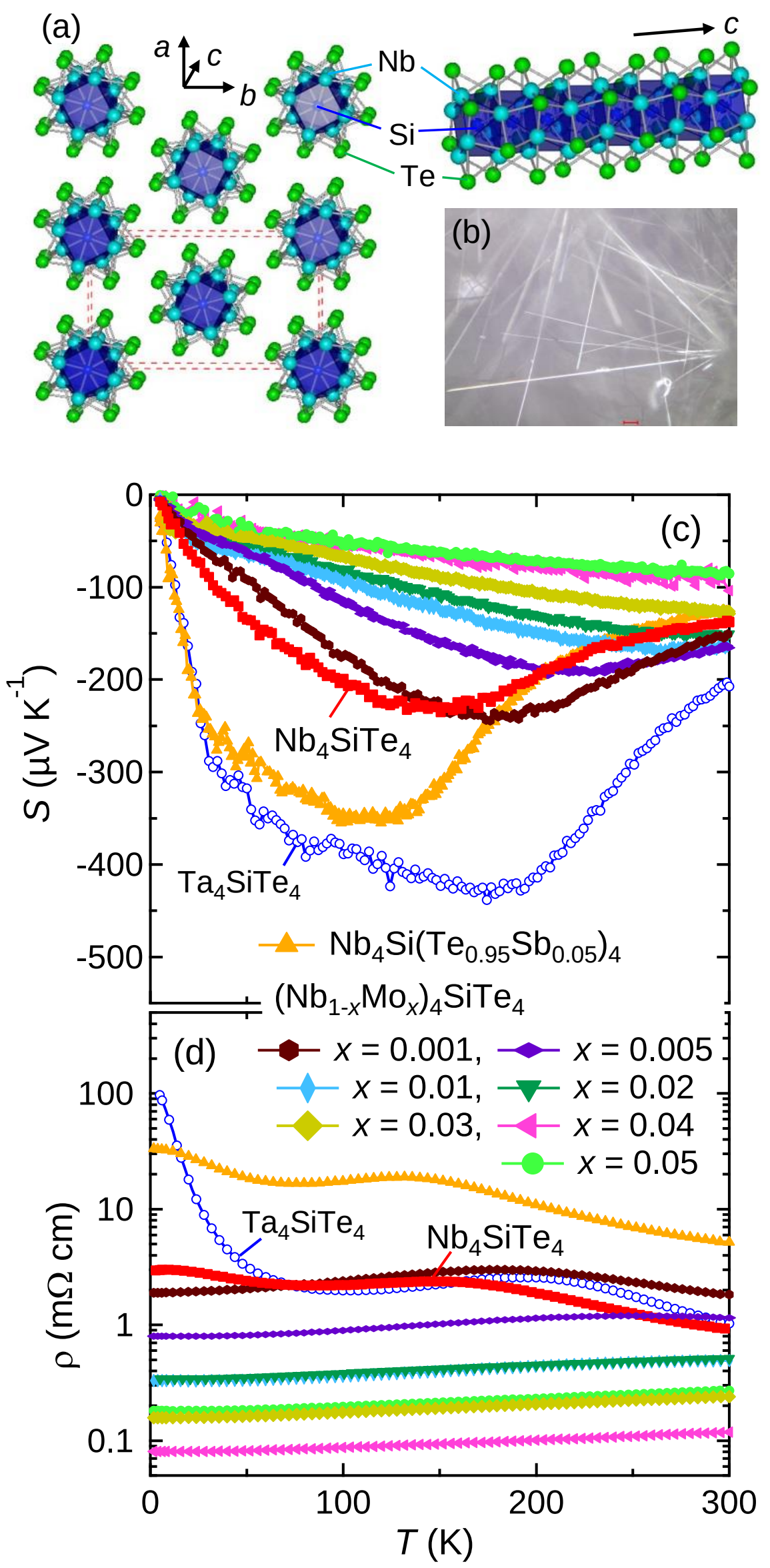

FIG. 1. (a) Crystal structure of $Nb_4SiTe_4$. The orthorhombic unit cell is indicated by broken lines. (b) A stereomicroscopic image of whisker crystals of Mo-doped $Nb_4SiTe_4$. The bar in the image indicates 50 μm. Temperature dependences of (c) thermoelectric power and (d) electrical resistivity of the whisker crystals of $Nb_4SiTe_4$, $(Nb_{1-x}Mo_x)_4SiTe_4$ ($x \leq 0.05$), and $Nb_4Si(Te_{0.95}Sb_{0.05})_4$ measured along the $c$ axis. The data of $Ta_4SiTe_4$ are also shown as a reference.[1]

the practical level. This $P$ is comparable to those of Mo-doped $Ta_4SiTe_4$ with similar doping levels but is smaller than the maximum value in $Ta_4SiTe_4$. The difference of thermoelectric performance of the Nb and Ta compounds might be understood by using first principles calculation. We also demonstrated that the $Nb_4SiTe_4$-$Ta_4SiTe_4$ solid solution also shows a large $P$ comparable to those of the end members, suggesting that $Nb_4SiTe_4$ can be utilized as a sister compound of $Ta_4SiTe_4$, as in the case of $Sb_2Te_3$ and $Bi_2Se_3$ for $Bi_2Te_3$.

The whisker crystals of $Nb_4SiTe_4$, $(Nb_{1-x}Mo_x)_4SiTe_4$ ($x \leq 0.05$), $Nb_4Si(Te_{0.95}Sb_{0.05})_4$, and $(Ta_{0.5}Nb_{0.5})_4SiTe_4$ were synthesized by crystal growth in a vapor phase. A stoichiometric amount of elemental powders and 100% excess of Si powder were mixed and sealed in an evacuated quartz tube with 20 mg of $TeCl_4$ powder. The tube was heated to and kept at 873 K for 24 h, 1423 K for 96 h, and then furnace cooled to room temperature. The typical size of the whisker crystals is several mm long and several μm in diameter, as shown in Fig. 1(b). A sintered sample of $Nb_4SiTe_4$ was prepared by a solid-state reaction method with the same procedure for $Ta_4SiTe_4$.[1] Sample characterization was performed by powder X-ray diffraction analysis for pulverized whisker crystals with Cu Kα radiation at room temperature using a RINT-2100 diffractometer (RIGAKU). We confirmed that the series of whisker crystals are a single phase. In this study, nominal compositions are used to represent the chemical compositions of the whisker crystals. The electrical resistivity and thermoelectric power measurements between 5 and 300 K were performed using a PPMS (Quantum Design). Thermal conductivity was measured by a standard four-contact method. First principles calculations for $Nb_4SiTe_4$ were performed using the WIEN2k code.[7] The Perdew-Burke-Ernzerhof generalized gradient approximation (PBE-GGA) is used for the exchange-correlation potential.[8] The self-consistent calculation is done for 9 × 5 × 20 $k$-mesh. The numbers of atoms contained in the unit cell are 16, 4, and 16 for Nb, Si, and Te, respectively. The radii of the muffin-tin spheres are $R_{MT}(Nb) = R_{MT}(Te) = 2.5a_0$ and $R_{MT}(Si) = 2.11a_0$, where $a_0$ is the Bohr radius ($R_{MT}(Ta) = R_{MT}(Te) = 2.5a_0$ and $R_{MT}(Si) = 1.99a_0$ in the calculation of $Ta_4SiTe_4$). We also set $R_{MT}K_{max} = 7.0$, $G_{max} = 12$, and core separation energy as –6.0 Ry. Experimental structural parameters are used for the calculations.

Figures 1(c) and 1(d) show the temperature dependences of $S$ and ρ of whisker crystals of $Nb_4SiTe_4$, $(Nb_{1-x}Mo_x)_4SiTe_4$ ($x \leq 0.05$), and $Nb_4Si(Te_{0.95}Sb_{0.05})_4$ measured along the $c$ axis, respectively. The $S$ of undoped $Nb_4SiTe_4$ is negative, indicating that the electron carriers are dominant. The $|S|$ of undoped $Nb_4SiTe_4$ is 150 μV $K^{-1}$ at 300 K and gradually increases with decreasing temperature, reaching a maximum value of 220 μV $K^{-1}$ at ~150 K, comparable to the $|S|$ of the $Bi_2Te_3$-based material at room temperature. Below 150 K, the $|S|$ decreases toward zero with decreasing temperature. As seen in Fig. 1(c), $|S|$ of $Nb_4SiTe_4$ is about half of that of $Ta_4SiTe_4$, although the thermoelectric powers of $Nb_4SiTe_4$ and $Ta_4SiTe_4$ have the same sign and similar temperature dependences. The ρ of the undoped $Nb_4SiTe_4$ is 1 mΩ cm at 300 K and increases to 2 mΩ cm at ~150 K with decreasing temperature, followed by a weak decrease below this temperature. Below ~70 K, ρ gradually increases again. As seen in Fig. 1(d), the values of ρ of $Nb_4SiTe_4$ and $Ta_4SiTe_4$ are almost the same above ~70 K, while they are significantly different below this temperature, where the increase of ρ of $Nb_4SiTe_4$ is much weaker than that of $Ta_4SiTe_4$.[1,5,6]

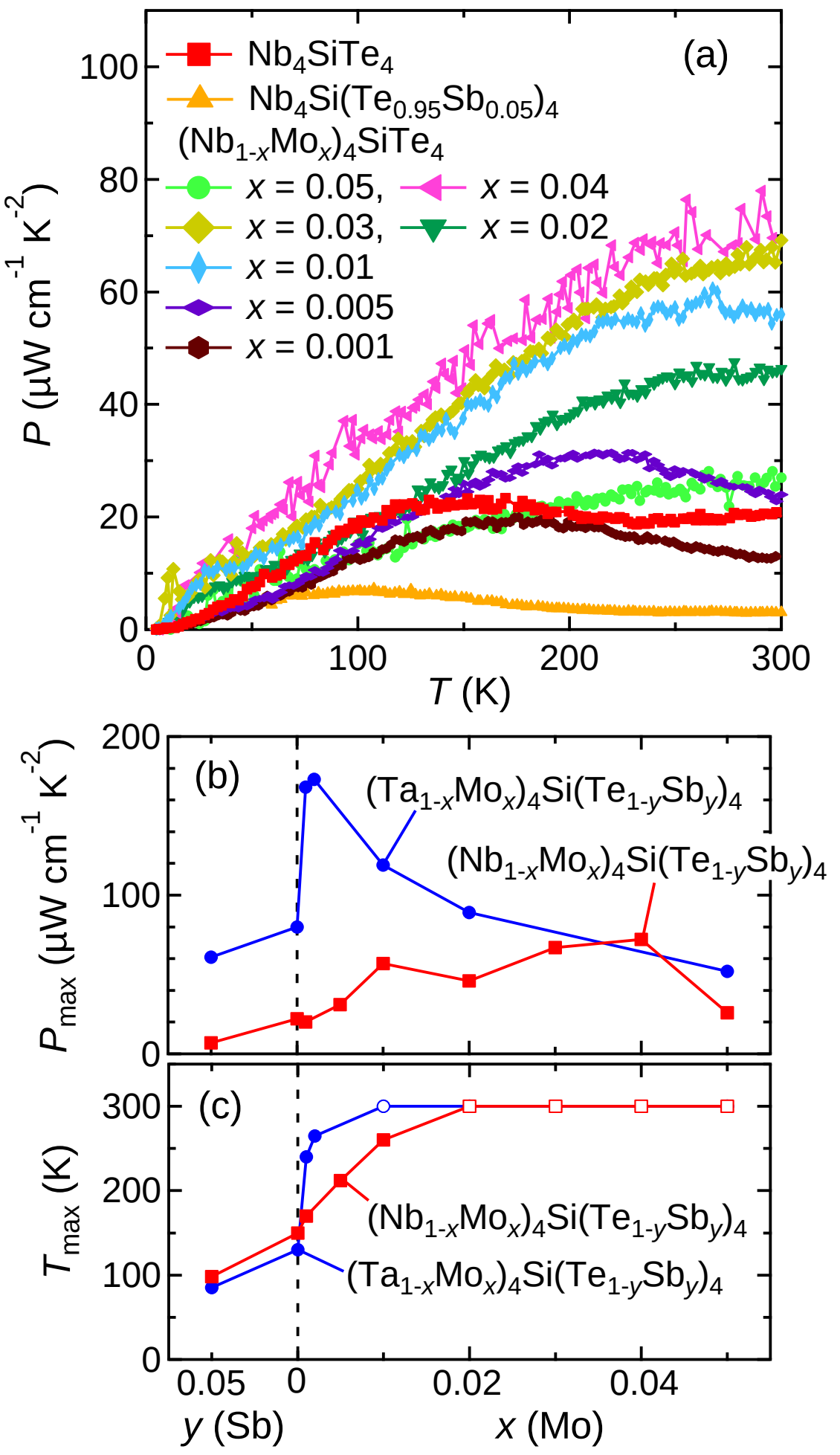

FIG. 2. (a) Temperature dependence of power factor of the whisker crystals of $Nb_4SiTe_4$, $(Nb_{1-x}Mo_x)_4SiTe_4$ ($x \leq 0.05$), and $Nb_4Si(Te_{0.95}Sb_{0.05})_4$ along the *c* axis. (b) Maximum power factor $P_{max}$ below 300 K. (c) Optimum temperature $T_{max}$, at which the power factor shows a maximum value $P_{max}$, below 300 K. The open symbols indicate $T_{max}$ = 300 K.

The thermoelectric powers of a series of $(Nb_{1-x}Mo_x)_4SiTe_4$ whisker crystals show a systematic change with Mo content. All measured samples have negative *S* values, as shown in Fig. 1(c), indicative of *n*-type. The peak temperature, where *S* shows a maximum value, increases with increasing Mo content from 140 K in the undoped sample. The samples with $x \geq 0.02$ show $d|S|/dT > 0$ in all temperature region below room temperature. $|S|$ systematically decreases with increasing *x*, except that the maximum value of $|S|$ of the $x = 0.001$ sample is slightly larger than that of the undoped one, consistent with the fact that Mo substitution to the Ta site is electron doping, which is expected to increase the electron carriers. The $x = 0.02$ and 0.05 samples show $|S| = 150$ and 80 μV $K^{-1}$ at 300 K, respectively, which are almost the same as those of $(Ta_{1-x}Mo_x)_4SiTe_4$ with the same *x* values.[1] Mo doping tends to decrease ρ, as shown in Fig. 1(d), reflecting the increase of electron carriers by electron doping. The ρ of the $x = 0.001$ sample shows a broad peak at ~180 K, larger than that of the undoped one above 70 K. The $x \geq 0.005$ samples show $d\rho/dT > 0$ in all temperature region below room temperature, suggestive of the presence of metallic conduction in them. The ρ of the $x = 0.05$ sample, which is most heavily doped in this study, is larger than those of $x = 0.03$ and 0.04, which might be due to the larger atomic disorder in the $x = 0.05$ sample caused by the Mo doping.

$Nb_4Si(Te_{0.95}Sb_{0.05})_4$ whisker crystal also shows negative *S* below 300 K, as shown in Fig. 1(c). Its $|S|$ and ρ are larger than those of the undoped one, respectively, probably reflecting the decrease of electron carriers by hole doping. The $|S|$ of $Nb_4Si(Te_{0.95}Sb_{0.05})_4$ reaches 350 μV $K^{-1}$ at 100-130 K, which is much larger than that of $Nb_4SiTe_4$ and comparable to that of $Ta_4SiTe_4$. The ρ of the $Nb_4Si(Te_{0.95}Sb_{0.05})_4$ sample is about an order of magnitude larger than that of $Nb_4SiTe_4$, although their temperature dependences are similar.

Figure 2(a) shows the power factor $P = S^2/\rho$ of the whisker crystals of $Nb_4SiTe_4$, $(Nb_{1-x}Mo_x)_4SiTe_4$ ($x \leq 0.05$), and $Nb_4Si(Te_{0.95}Sb_{0.05})_4$. The undoped $Nb_4SiTe_4$ shows $P \sim 20$ μW $cm^{-1}$ $K^{-2}$ between 100 and 300 K and a maximum value of $P_{max}$ = 22 μW $cm^{-1}$ $K^{-2}$ at the optimum temperature of $T_{max}$ = 150 K. The *P* values tend to increase with increasing *x*, reaching 70 μW $cm^{-1}$ $K^{-2}$ above 230 K at $x = 0.04$. This power factor is smaller than $P_{max}$ = 170 μW $cm^{-1}$ $K^{-2}$ for the Mo-doped $Ta_4SiTe_4$, as shown in Fig. 2(b), but is larger than $P_{max}$ = 35 μW $cm^{-1}$ $K^{-2}$ of the $Bi_2Te_3$-based materials,[9,10] indicating that $Nb_4SiTe_4$ is a strong candidate for a low-temperature thermoelectric material same as $Ta_4SiTe_4$. As seen in Fig. 2(b), the *P* values of the Nb compounds are optimized at $x$ = 0.03–0.04, which is significantly larger than $x < 0.01$ for the Ta compounds. Moreover, the *x* dependence of $T_{max}$ of the Nb compounds is more gradual than that of the Ta ones, as shown in Fig. 2(c). These results suggest that the thermoelectric performances of the Nb compounds are less sensitive to the chemical doping, i.e., more easily optimized than those of the Ta compounds.

We note the dimensionless figure of merit $ZT = PT/(\kappa_{el} + \kappa_{lat})$ of Mo-doped $Nb_4SiTe_4$, where $\kappa_{el}$ and $\kappa_{lat}$ are the conduction electron and phonon contributions of thermal conductivity, respectively. *ZT* is directly related to the thermoelectric energy conversion efficiency. Although thermal conductivity of the whisker crystals prepared in this study cannot be measured due to their very thin shape, we estimated the upper limit of *ZT* of them by using $\kappa_{el}$ of the whisker crystals, obtained by applying Wiedemann-Franz law to the ρ data shown in Fig. 1(d), and $\kappa_{lat}$ of the $Nb_4SiTe_4$ sintered sample. As shown in Fig. 3, thermal conductivity κ of a $Nb_4SiTe_4$ sintered sample is 11 mW $cm^{-1}$ $K^{-1}$ at room temperature, almost all of which is the phonon contribution, considering ρ of the sintered sample. The upper limit of *ZT* for $x$ = 0.03 at room temperature, where *P* is optimized as shown in Fig. 2(b), is estimated to be 0.5, which is smaller than 2.2 for $(Ta_{0.999}Mo_{0.001})_4SiTe_4$ at 250 K.[1]

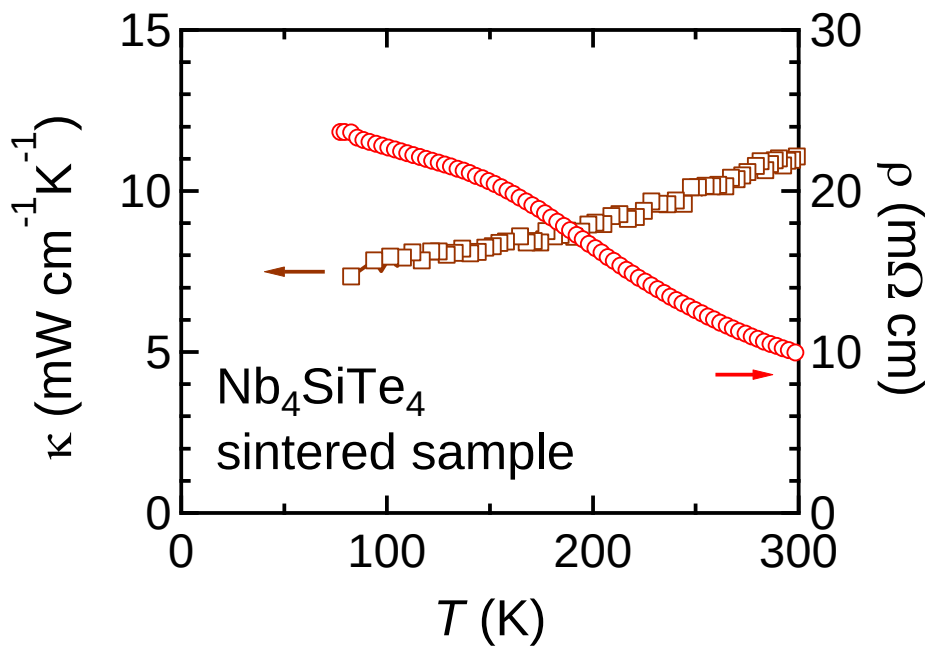

FIG. 3. Temperature dependences of thermal conductivity and electrical resistivity of a $Nb_4SiTe_4$ sintered sample.

Thus, the whisker crystals of the Nb compounds were found to show the large *P* and upper limit of *ZT*, although they are smaller than the maximum values for the Ta compounds. The difference of *P* values between the Nb and Ta compounds depends on the kind and content of the dopant ions. In the undoped and Sb-doped cases, where electron carrier density is expected to be lower than the Mo-doped case, the *P* of the Nb compounds are much smaller than those of Ta ones, as shown in Fig. 2(b). In contrast, heavily Mo-doped $Nb_4SiTe_4$ shows large *P* comparable to that of Mo-doped $Ta_4SiTe_4$. Both $(Nb_{1-x}Mo_x)_4SiTe_4$ and $(Ta_{1-x}Mo_x)_4SiTe_4$ with $x \geq 0.01$ exhibit $d\rho/dT > 0$ and $dS/dT > 0$ below room temperature, different from the undoped $Nb_4SiTe_4$ and $Ta_4SiTe_4$, probably reflecting the larger electron carrier density.[1] They also show similar |*S*| values, giving rise to similar *P* values, although there is some difference due to the different ρ values.

Here, we discuss the doping dependences of the thermoelectric properties of the Nb- and Ta-based whisker crystals in the light of their electronic structures. Figures 4(a) and 4(b) show the first principles calculation results of $Nb_4SiTe_4$ with and without spin-orbit coupling, respectively. The electronic structures of $Nb_4SiTe_4$ are similar to those of $Ta_4SiTe_4$, reflecting the same crystal structure and electron configuration.[1] The band dispersions along the $k_x$ and $k_y$ directions are almost flat and much weaker than that along the $k_z$ direction. This strongly one-dimensional electronic structure would be one of the important factors to realize high thermoelectric performances in $Nb_4SiTe_4$ and $Ta_4SiTe_4$.[11,12] We also confirmed that the Mo doping to $Nb_4SiTe_4$ results in the upward shift of the chemical potential (5% Mo doping gives ~0.2 eV shift) without a significant change of the band structure by the band calculations using the virtual crystal approximation.[13]

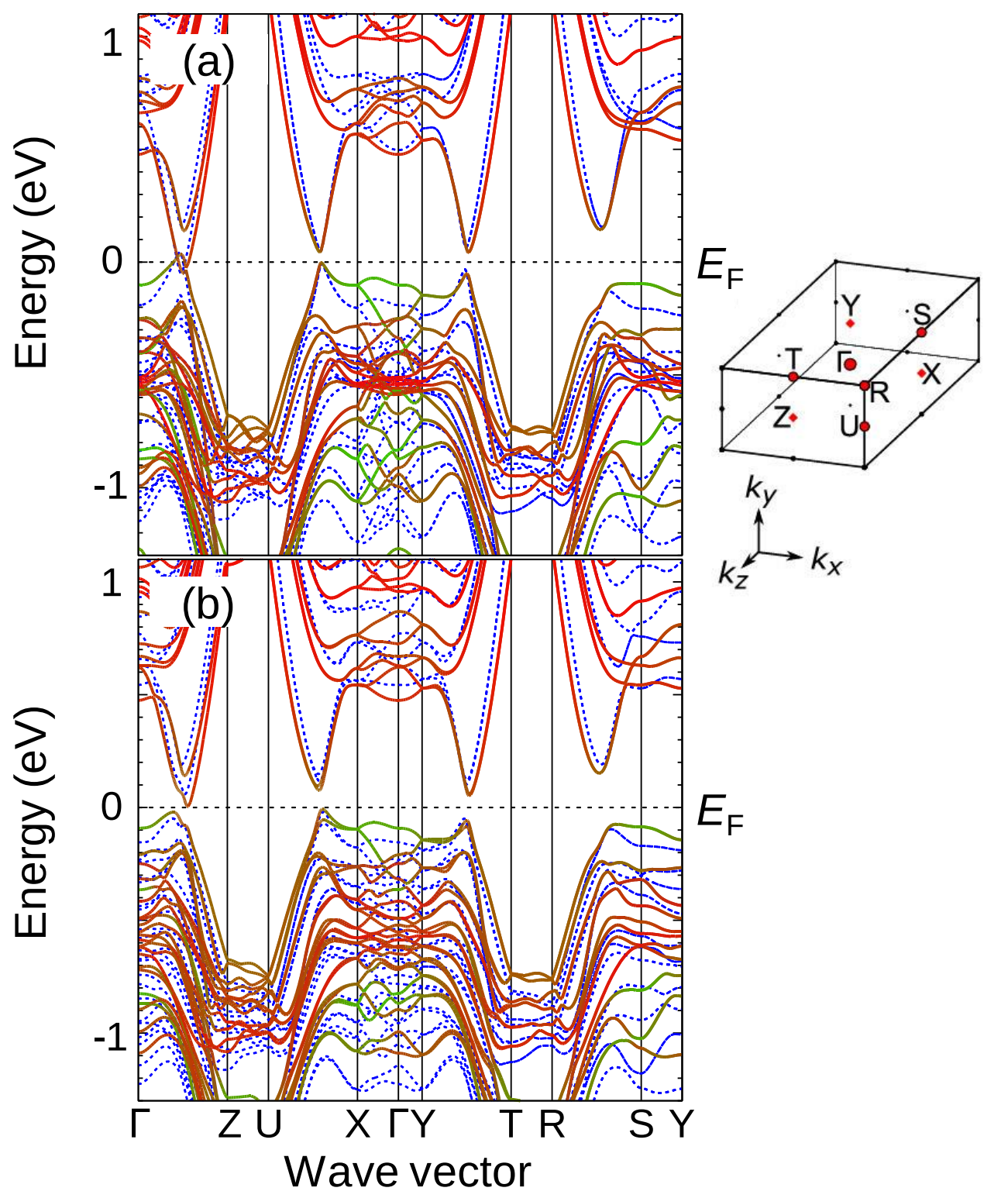

FIG. 4. Electronic structures of $Nb_4SiTe_4$ without (a) and with (b) spin-orbit coupling. Red and green represent the contributions of the *d*-orbital of Nb and *p*-orbital of Te, respectively. The broken lines indicate those of $Ta_4SiTe_4$.[1] The Fermi level is set to 0 eV. The first Brillouin zone is shown in the right panel of (a).

The difference of thermoelectric performances between the Nb and Ta compounds seems to be caused by a much smaller spin-orbit gap in $Nb_4SiTe_4$. When the spin-orbit coupling is switched off, the energy bands cross on the Γ–Z line at around $E_F$ in both $Nb_4SiTe_4$ and $Ta_4SiTe_4$. In $Ta_4SiTe_4$, the strong spin-orbit coupling of Ta 5*d* orbitals gives rise to an energy gap of $\Delta$ = 0.1 eV near the band crossing points.[1] In contrast, the spin-orbit gap of $\Delta \sim 0.02$ eV in $Nb_4SiTe_4$ is much smaller than that of $Ta_4SiTe_4$. These $\Delta$ values are probably related to the different behaviors of ρ of the undoped $Nb_4SiTe_4$ and $Ta_4SiTe_4$ below 50 K mentioned above. We believe that the size of $\Delta$ also plays an important role in the thermoelectric performances, such as the significantly different $P_{max}$ between the Nb and Ta compounds in the left side of Fig. 4(b), where the electron carrier densities are lower than those of the right side. In contrast, the *S* and ρ of the heavily Mo-doped $Nb_4SiTe_4$ and $Ta_4SiTe_4$ show simple metallic behaviors, suggesting that the $E_F$ of them is located well above the bottom of the conduction band due to the electron doping. In this case, fine structures of the energy bands at around $E_F$, such as the size of $\Delta$, seem to have little effect on the thermoelectric properties and result in similar thermoelectric performances in the Nb and Ta compounds.

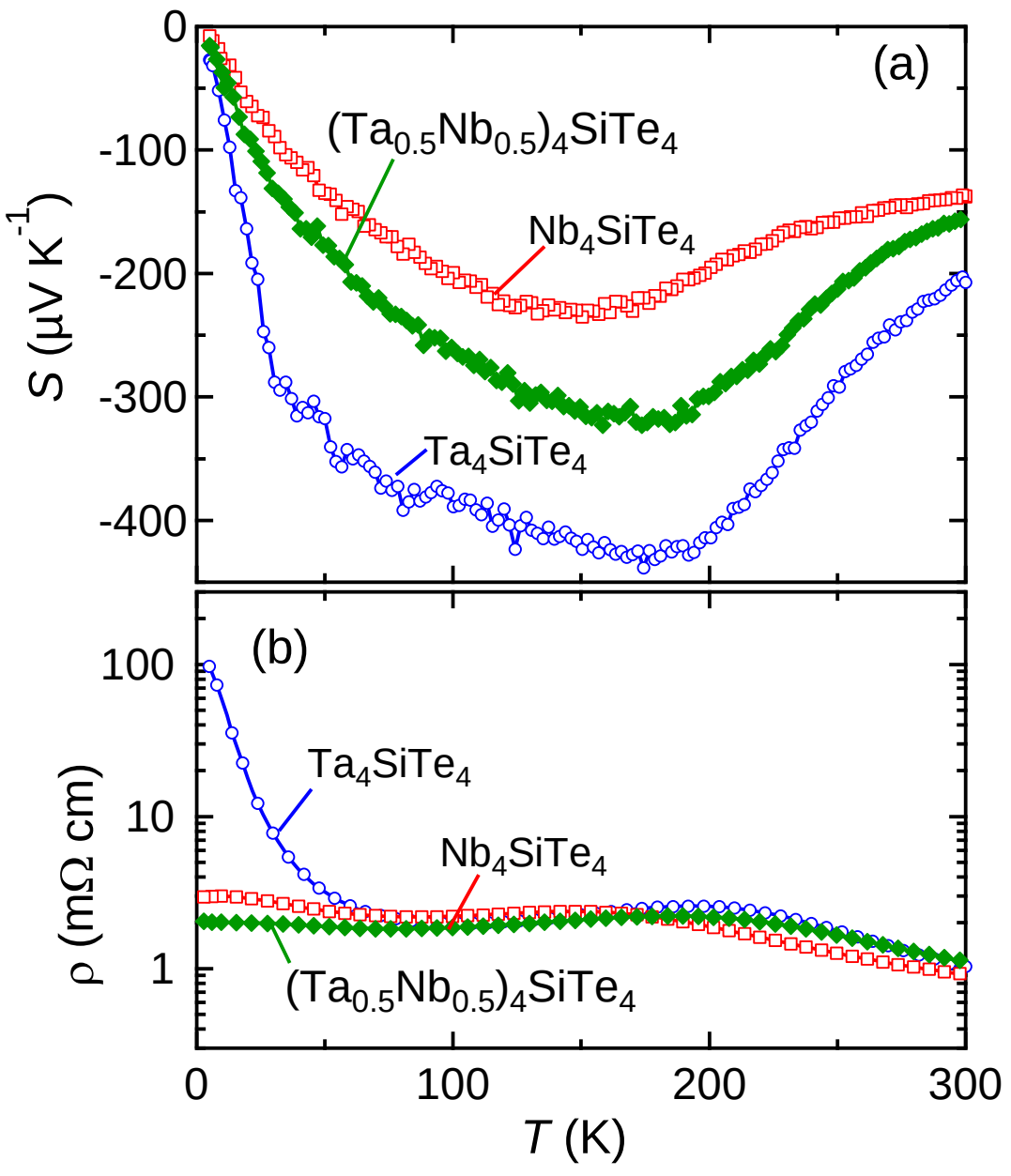

FIG. 5. Temperature dependences of thermoelectric power (a) and electrical resistivity (b) of the whisker crystals of $(Ta_{0.5}Nb_{0.5})_4SiTe_4$, $Nb_4SiTe_4$, and $Ta_4SiTe_4$ measured along the $c$ axis.

Finally we report the thermoelectric properties of a $Nb_4SiTe_4$-$Ta_4SiTe_4$ solid solution sample. Figures 5(a) and 5(b) show the temperature dependences of $S$ and ρ of $(Ta_{0.5}Nb_{0.5})_4SiTe_4$, $Nb_4SiTe_4$, and $Ta_4SiTe_4$ whisker crystals, respectively. The $S$ of $(Ta_{0.5}Nb_{0.5})_4SiTe_4$ is negative and located between those of $Nb_4SiTe_4$ and $Ta_4SiTe_4$. The $|S|$ of $(Ta_{0.5}Nb_{0.5})_4SiTe_4$ shows a maximum value of 310 μV $K^{-1}$ at 180 K, indicating that the solid-solution sample has a sufficient $|S|$ as a thermoelectric material. The ρ of the solid-solution sample is almost the same as those of $Nb_4SiTe_4$ and $Ta_4SiTe_4$ above ~100 K. Below this temperature, the ρ of the solid-solution sample is much smaller than that of $Ta_4SiTe_4$, but is comparable to that of $Nb_4SiTe_4$. This may reflect the presence of a significant contribution of Nb 4$d$ orbitals near the $E_F$ in the solid solution samples, which is expected to give rise to a very small spin-orbit gap same as in $Nb_4SiTe_4$. The $P_{max}$ of the solid solution sample is 47 μW $cm^{-1}$ $K^{-2}$ at $T_{max}$ = 160 K, which exceeds the practical level, suggesting that the combination of $Nb_4SiTe_4$ and $Ta_4SiTe_4$ is promising as a practical thermoelectric material, same as the $Bi_2Te_3$-$Sb_2Te_3$ and Si-Ge alloys.

In conclusion, we have studied the thermoelectric properties of whisker crystals of $Nb_4SiTe_4$, a 4$d$ analogue of $Ta_4SiTe_4$, and its substituted compounds and found that Mo-doped $Nb_4SiTe_4$ shows high thermoelectric performances at low temperatures. The power factor of the undoped $Nb_4SiTe_4$ is smaller than the maximum value of $Ta_4SiTe_4$, while those of $(Nb_{1-x}Mo_x)_4SiTe_4$ with $x \geq 0.01$ are very large, as represented by 70 μW $cm^{-1}$ $K^{-2}$ for $x$ = 0.03–0.04 at 230-300 K, comparable to those in the Ta compounds. The electronic structure of $Nb_4SiTe_4$ is similar to that of $Ta_4SiTe_4$ in that a small spin-orbit gap opens in the strongly one-dimensional electronic bands, probably yielding the high thermoelectric performances in both systems. Alternatively, the difference of thermoelectric performances between Nb and Ta compounds with low electron carrier density might be related to the significantly smaller spin-orbit gap in $Nb_4SiTe_4$ than that in $Ta_4SiTe_4$, suggesting that the spin-orbit gap of ~0.1 eV in $Ta_4SiTe_4$ plays an important role in the colossal power factor in the lightly Mo doped $Ta_4SiTe_4$. Moreover, we found that not only the Nb compounds but also the solid solution of the Nb and Ta compounds shows a high thermoelectric performance exceeding those of practical thermoelectric materials, indicating that the combination of $Nb_4SiTe_4$ and $Ta_4SiTe_4$ is promising to realize practical-level thermoelectric conversion at low temperatures.

## ACKNOWLEDGMENTS

We are grateful to Y. Yoshikawa for his help with experiments and A. Yamakage for helpful discussion. This work was partly supported by JSPS KAKENHI (Grant Nos. 16K13664, 16H03848, and 16H01072), the Murata Science Foundation, and the Tatematsu Foundation.

## REFERENCES

[1]T. Inohara, Y. Okamoto, Y. Yamakawa, A. Yamakage, and K. Takenaka, Appl. Phys. Lett. **110**, 183901 (2017).

[2]M. E. Badding and F. J. DiSalvo, Inorg. Chem. **29**, 3952 (1990).

[3]J. Li, R. Hoffmann, M. E. Badding, and F. J. DiSalvo, Inorg. Chem. **29**, 3943 (1990).

[4]R. Venkatasubramanian, E. Siivola, T. Colpitts, and B. O'Quinn, Nature **413**, 597 (2001).

[5]M. E. Badding, R. L. Gitzendanner, R. P. Ziebarth, and F. J. DiSalvo, Mat. Res. Bull. **29**, 327 (1994).

[6]K. Ahn, T. Hughbanks, K. D. D. Rathnayaka, and D. G. Naugle, Chem. Mater. **6**, 418 (1994).

[7]P. Blaha P., K. Schwarz, G. Madsen, D. Kvasnicka, and J. Luitz, *WIEN2k, an augmented plane wave + local orbitals program for calculating crystal properties* (Tech. Universität Wien, Vienna, 2001).

[8]J. P. Perdew, K. Burke, and M. Ernzerhof, Phys. Rev. Lett. **77**, 3865 (1996).

[9]G. D. Mahan, *Solid State Physics* (Academic Press, New York, USA, 1997) Vol. 51, pp. 81-157.

[10]L. R. Testardi, J. N. Bierly Jr., and F. J. Donahoe, J. Phys. Chem. Solid **23**, 1209 (1962).

[11]V. A. Greanya, W. C. Tonjes, R. Liu, C. G. Olson, D.-Y. Chung, and M. G. Kanatzidis, Phys. Rev. B **65**, 205123 (2002).

[12]P. Larson, S. D. Mahanti, D.-Y. Chung, and M. G. Kanatzidis, Phys. Rev. B **65**, 045205 (2002).

[13]L. Nordheim, Ann. Phys. (Leipzig) **9**, 607 (1931).